\documentclass{JAC2000}

\usepackage{graphicx}

\begin{document}
\title{Design Studies for a High Current Bunching System 
for CLIC Test Facility (CTF3) Drive Beam}

\author{Y. Thiery, J. Gao, and J. Le Duff, LAL, Orsay, FRANCE}

\maketitle

\begin{abstract} 
A bunching system is proposed for the initial stage (bunch spacing is 10 cm)
of 
CTF3 which consists of one (two) 3 GHz prebuncher(s) and
one 3GHz travelling wave (TW) 
buncher with variable phase velocities ($\beta =0.75$ and 1) working at 
$2\pi \over 3$ mode.   
Since the average macropulse beam current (3.5 A) 
at the exit of the TW buncher 
is rather high, inside the TW buncher one has to
take the beam loading effect into consideration.
By using PARMELA, it is shown numerically
that
the bunching system can provide the bunches which properties satisfy the design requirement
of CTF3. 
The dimensions of the cavities in the two phase velocity regions are proposed.
The transient beam loading effect and the 
multibunch transverse instabilities are studied numerically.
It is concluded that higher order mode (HOM) couplers should be installed
in the TW buncher with the loaded quality factor of the 
dipole mode lower than 80.
\end{abstract}

\section{Introduction}
CLIC is a two beam accelerator (TBA) based e$^+$e$^-$ linear collider. 
The recently proposed
CLIC drive beam scheme \cite{ruth} makes CLIC more interesting. To demonstrate
the feasibility of the new drive beam scheme and to test other technical 
aspects, CTF3 \cite{clic}
has been proposed as 
a natural successor of the existing CTF2 to which LAL has actively 
collaborated and contributed in the past years \cite{gao1}\cite{gao2}. 
In this paper we will restrict ourselves to the study of the bunching
system 
for CTF3 which is a new subject of collaboration between LAL and
CLIC group of CERN.
\par
The bunching system under study consists of 
a 140 KV DC gun, one (two) prebuncher(s) of 3 GHz for the initial stage
and one TW buncher of 3 GHz. 
In the following sections we will discuss 
the design
of the travelling wave buncher considering the beam loading effect,
the 
multibunch longitudinal and transverse beam dynamics in the 
TW buncher with the presence of long range wakefields, and
the numerical simulations of the proposed bunching system by using PARMELA. 
\par
\section{Beam loading effect}
We start with the power diffusion equation in a linac
\begin{equation}
{dP(z)\over dz}=-2\alpha P(z)-IE(z)
\label{eq:1}
\end{equation} 
where $\alpha =\omega /(2Qv_g)$, $v_g$ is the group velocity,
$Q$ is the quality factor, $\omega$ is the angular working frequency,
$P(z)$ is the power flow inside the structure,
$E(z)$ is the amplitude of the synchronous accelerating field,
and $I$ is the average beam current during the rf pulse.
By using the initial condition $E(0)=E_0$, one gets:
\begin{equation}
E(z)=E_0\exp(-\alpha z)\sin(\phi)-IR_{sh}(1-\exp(-\alpha z))
\label{eq:6}
\end{equation} 
where $R_{sh}(z)$ is the shunt impedance of the accelerating mode,
$E_0=\sqrt{2\alpha P_0R_{sh}}$ and $P_0$ is the input power from 
the rf source,
$R_{sh}$ and $\alpha$ are kept constant within the accelerating
structure, 
$\phi$ is the phase of the accelerating field
($\phi =\pi /2$ corresponds to the maximum acceleration), and this
expression was first obtained by A.J. Lichtenberg \cite{licht}.
\par
\begin{figure}[t]
\vspace*{10.5cm}
\includegraphics{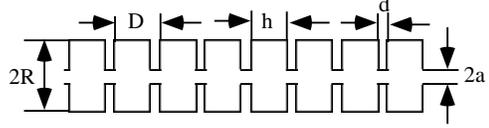}
\vspace{-9cm}
\caption{Disk-loaded accelerating structure \label{fig:1q}}
\end{figure}

To have good energy transfer efficiency and energy gain at the same time,
one requires (fully beam loading condition):
\begin{equation}
E(L)=0
\label{eq:8}
\end{equation}
where $L$ is the length of the accelerating section.
When the ohmic losses on the structure wall are small the fully beam loaded condition
results in
\begin{equation}
P_0={IE_0L\over 2}
\label{eq:9}
\end{equation} 
For a given $E_0$, $I$, and $L$, one gets $P_0$ from eq. \ref{eq:9} and one
can determine the geometry of the disk-loaded structure by solving the
following equations:
\begin{equation}
{E_0^2\over P_0}={\omega R_{sh}\over v_gQ}={4k_{010}\over c (v_g/c)}
\label{eq:14}
\end{equation}
where $k_{010}$ and $v_g/c$ can be expressed analytically as 
\cite{gao3}\cite{gao4}:
\begin{equation}
k_{010}={2 J_0^2\left({u_{01}\over R}a\right)\sin^2\left({u_{01}h\over 2R}\right)\over \epsilon_0\pi hDJ_{1}^2(u_{01})u_{01}^2}
\label{eq:10}
\end{equation} 
\begin{equation}
{v_g\over c}={\omega K_eD\sin(\theta_0)\over 2c}
\label{eq:12}
\end{equation} 
\begin{equation}
K_e={4a^3\over 3\pi hR^2J_1^2(u_{01})}\exp({-\alpha_ed})
\label{eq:11}
\end{equation} 
\begin{equation}
\alpha_e=\left(\left({2.405 /a}\right)^2-\left({2\pi /\lambda}\right)^2\right)^{1/2}
\label{eq:13}
\end{equation}
where $a$, $h$, $R$, $D$, and $d$ are defined in 
Fig. \ref{fig:1q}, $u_{01}=2.405$, $\lambda$ is the wavelength in free space,
and $\theta_0$ is the phase shift per cell.
Now we make a rough design for the TW buncher working at 3 GHz 
(R$\approx 0.04$ m) and 2$\pi$/3 mode. If one takes $E_0=10$ MV/m,
$I=4$ A, $L=0.8$ m, one finds $P_0=16$ MW and the structure dimensions
given in Table 1. The structure has two sections with phase velocities
$\beta =0.75$ and $\beta =1$, respectively. The number of the cells of
$\beta =0.75$ is four which has been determined by the beam dynamics 
simulations of PARMELA. 
\begin{table}[t]
\begin{center}
\begin{tabular}{|l|l|l|l|l|}
\hline
Cell type&D (m)
&h (m)&a (m)&$v_g/c$\\
\hline
\hline
$\beta =0.75$& 0.025
&0.0137 &0.0146 &0.031\\
\hline
$\beta =1$&0.033 
&0.02 &0.016 &0.029\\
\hline
\end{tabular}
\end{center}
\vskip -0.3 true cm
\caption{summary of TW buncher dimension}
\label{tab:1}
\end{table}
\begin{figure}[htb]
\centering
\includegraphics*[width=80mm]{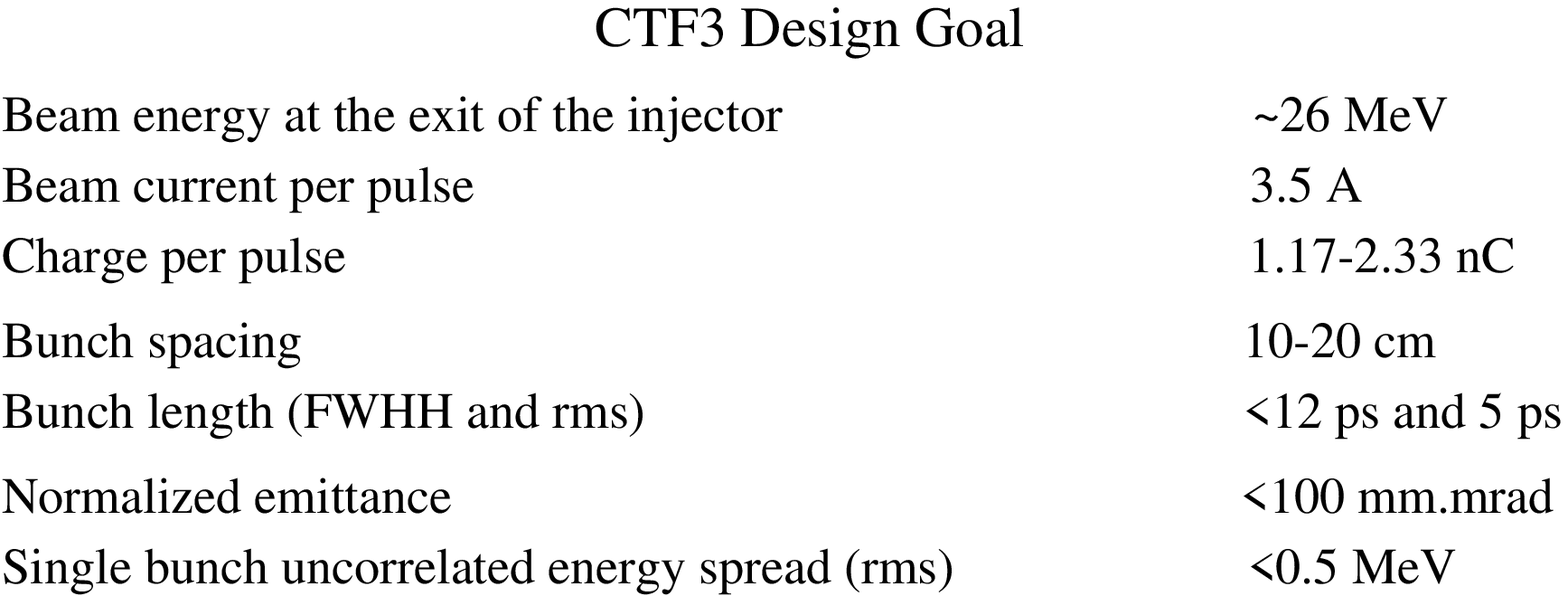}
\caption{The design goal.}
\label{f1}
\end{figure}  
\begin{figure}[htb]
\centering
\includegraphics*[width=80mm]{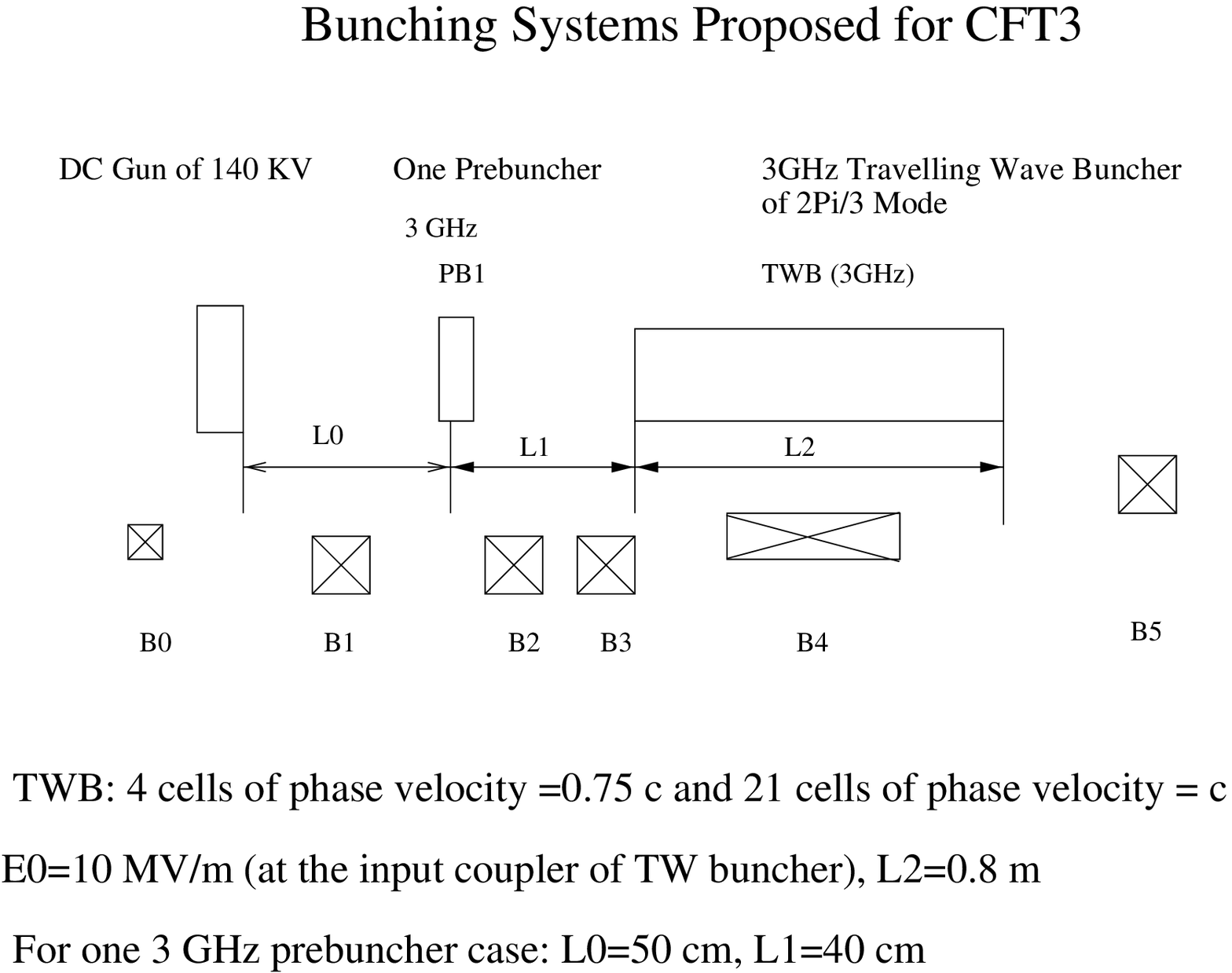}
\vskip -0.3 true cm
\caption{The schematic layout of the bunching system.}
\label{f2}
\end{figure}
\par
\section{Beam dynamics}
\begin{figure}[t]
\vskip 0. true cm
\vspace{10cm}
\includegraphics{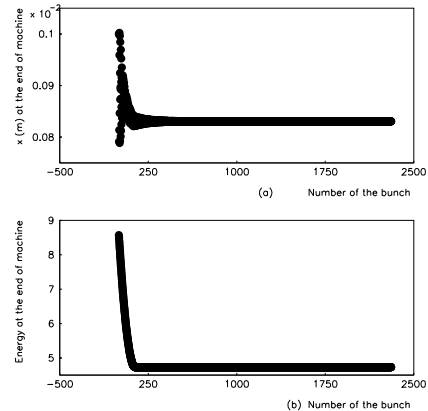}
\vskip -4.5 true cm
\caption{The bunch separation is 20 cm (final phase of CTF3), 
and the TW buncher is damped with
$Q_{L,110}=80$. At the exit of the TW buncher:
(a) The transverse motion of a bunch train with an initial offset
of 1 mm. (b) The energy gain of the bunch train.
\label{f3}}
\end{figure}

\begin{figure*}[t]
\centering
\includegraphics*[width=130mm]{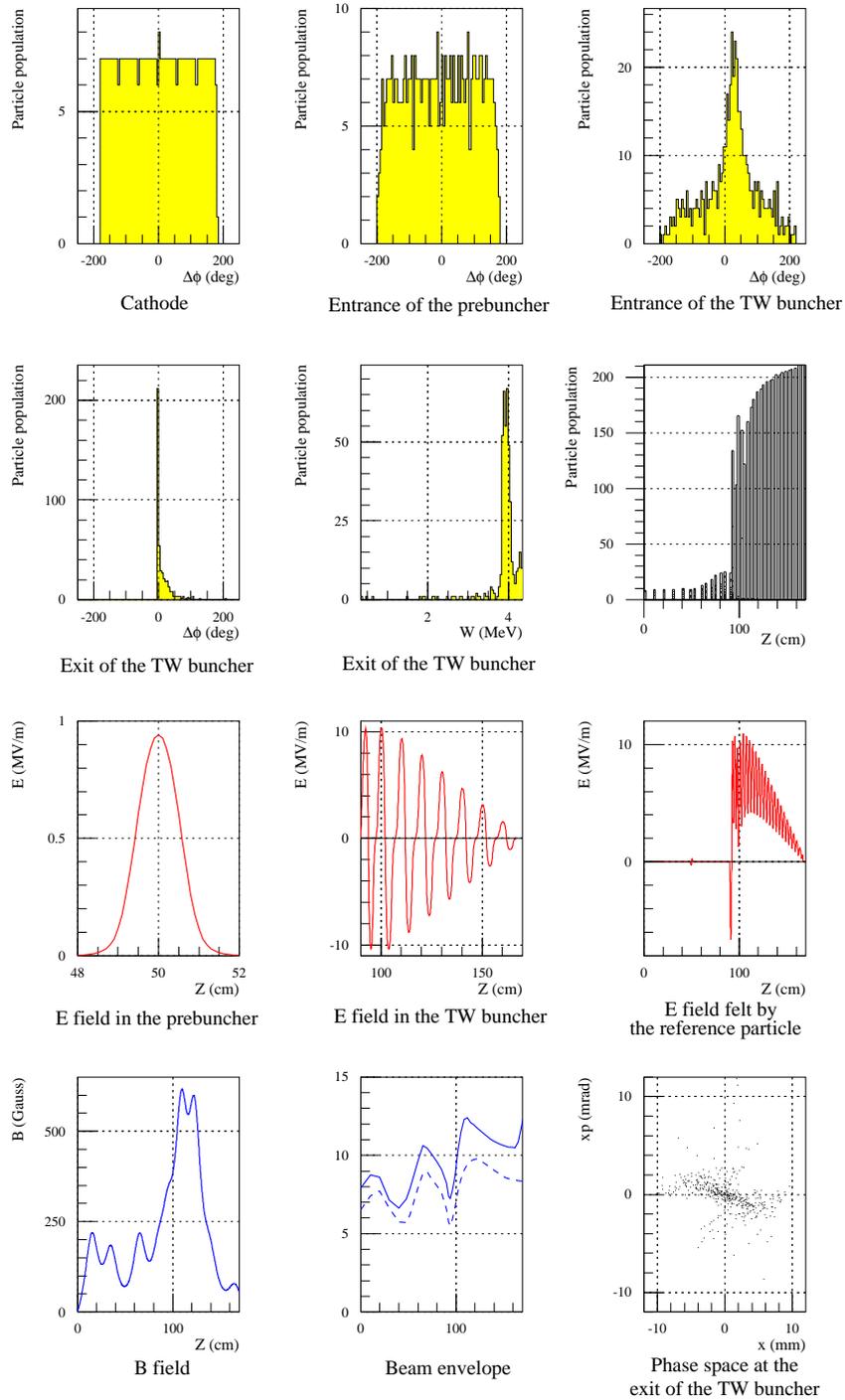}
\caption{PARMELA simulation results.}
\label{f4}
\end{figure*}

The design goal and the layout of the bunching system are
shown in Fig. \ref{f1},
and in Fig. \ref{f2} \cite{loui}\cite{hans}.
A multibunch beam dynamics study shows that 
for the initial stage of CTF3 there is no need to damp the dipole modes in the TW buncher; however,
for the final phase
(bunch spacing is 20 cm) 
the dipole modes in the TW buncher have to be damped to $Q_{L,110}\leq 80$ as shown in 
Fig. \ref{f3}. By using PARMELA one gets the bunched beam parameters
at the exit of the TW buncher. Now we only show 
the simulation results in Fig. \ref{f4}
for the single 3 GHz prebuncher case. 
The DC current coming from the cathode is
7 A and during 2$\pi$, 500 particles have been tracked.   
For the observing
window of 20 degree centered around the bunch current peak, one gets
322 particles transmitted at the exit of the TW buncher, the normalized rms emittance is 51 mm.mrad, the rms energy
spread is 0.12 MeV, the energy of the reference particle is 3.8 MeV, and the
longitudinal rms bunch length is 4.4 ps. In future simulations two
fully beam loaded structures will be added after the TW buncher to push
the beam energy up to about 26 MeV.
\par 

\section{Conclusion}
We have given a preliminary design for the TW buncher. By
using PARMELA the bunching system consisting one (two) prebuncher(s) and 
a TW buncher has been simulated and the results are satisfactory. More simulations will be done to determine how many prebunchers are to be used and to
include two accelerating sections to accelerate the beam to about 26 MeV. Multibunch beam dynamic simulation results show that the
loaded quality factors of the dipole modes in the 
TW buncher have to be lower than 80.  
\par
\section{Acknowledgement}
We thank L. Rinolfi and E. Jensen for providing useful information
and discussions.
\par

\end{document}